\documentclass[
aps,preprintnumbers,amsmath,showpacs,amssymb,prl,nofootinbib,superscriptaddress,reprint]{revtex4-1}

\usepackage[dvipdfmx]{graphicx}
\usepackage{dcolumn}

\usepackage[colorlinks=true,allcolors=blue]{hyperref}

\begin{document}


\def\aa{{\dot \a}}
\def\bb{{\dot \b}}
\def\ss{{\bar \s}}
\def\hh{{\bar \h}}
\def\CA{{\cal A}}
\def\CB{{\cal B}}
\def\CC{{\cal C}}
\def\CD{{\cal D}}
\def\CE{{\cal E}}
\def\CG{{\cal G}}
\def\CH{{\cal H}}
\def\CI{{\cal I}}
\def\CK{{\cal K}}
\def\CL{{\cal L}}
\def\CR{{\cal R}}
\def\CM{{\cal M}}
\def\CN{{\cal N}}
\def\CO{{\cal O}}
\def\CP{{\cal P}}
\def\CQ{{\cal Q}}
\def\CS{{\cal S}}
\def\CT{{\cal T}}
\def\CW{{\cal W}}
\newcommand{\bH}{\mathbb{H}}
\newcommand{\bC}{\mathbb{C}}
\newcommand{\bR}{\mathbb{R}}
\newcommand{\bZ}{\mathbb{Z}}
\newcommand{\bS}{\mathbb{S}}
\def\Nequals#1{$\mathcal{N}{=}#1$}
\def\SU{\mathrm{SU}}
\def\Sp{\mathrm{Sp}}
\def\USp{\mathrm{USp}}
\def\U{\mathrm{U}}
\def\SO{\mathrm{SO}}
\def\O{\mathrm{O}}
\def\SL{\mathrm{SL}}
\def\tr{\mathop{\mathrm{tr}}\nolimits}
\def\diag{\mathop{\mathrm{diag}}\nolimits}
\def\rank{\mathop{\mathrm{rank}}\nolimits}
\def\sign{\mathop{\mathrm{sign}}\nolimits}
\def\Tr{\mathop{\mathrm{Tr}}\nolimits}
\def\II{I\hspace{-.1em}I}
\def\vev#1{\langle#1\rangle}
\def\cH{{\cal H}}
\def\ff{\mathfrak{f}}
\def\fg{\mathfrak{g}}
\def\CP{\mathbb{CP}}
\def\RP{\mathbb{RP}}
\def\sB{\mathsf{B}}
\def\sS{\mathsf{S}}

\newcommand{\Slash}[1]{{\ooalign{\hfil/\hfil\crcr$#1$}}}

\def\o{\over}
\newcommand{\gsim}{ \mathop{}_{\textstyle \sim}^{\textstyle >} }
\newcommand{\lsim}{ \mathop{}_{\textstyle \sim}^{\textstyle <} }
\newcommand{\bra}[1]{ \langle {#1} | }
\newcommand{\ket}[1]{ | {#1} \rangle }
\newcommand{\EV}{ {\rm eV} }
\newcommand{\KEV}{ {\rm keV} }
\newcommand{\MEV}{ {\rm MeV} }
\newcommand{\GEV}{ {\rm GeV} }
\newcommand{\TEV}{ {\rm TeV} }
\def\diag{\mathop{\rm diag}\nolimits}
\def\Spin{\mathop{\rm Spin}}
\def\SO{\mathop{\rm SO}}
\def\O{\mathop{\rm O}}
\def\SU{\mathop{\rm SU}}
\def\U{\mathrm{U}}
\def\Sp{\mathop{\rm Sp}}
\def\USp{\mathop{\rm USp}}
\def\SL{\mathop{\rm SL}}
\def\tr{\mathop{\rm tr}}
\def\rank{\mathop{\rm rank}}

\def\spin{\mathfrak{ spin}}
\def\so{\mathfrak{so}}
\def\o{\mathfrak{o}}
\def\su{\mathfrak{su}}
\def\u{\mathfrak{u}}
\def\sp{\mathfrak{sp}}
\def\usp{\mathfrak{usp}}
\def\sl{\mathfrak{sl}}
\def\e{\mathfrak{e}}

\newcommand*\Laplace{\mathop{}\!\mathbin\bigtriangleup}

\def\beq#1\eeq{\begin{align}#1\end{align}}
\def\alert#1{{\color{red}[#1]}}

\def\alph{\alpha}
\def\lambd{\lambda}

\def\4d{4D}
\def\2d{2D}
\def\1d{1D}


\preprint{IPMU19-0162}
\preprint{TU-1095}

\title{Confinement as Analytic Continuation Beyond Infinity}

\author{Masahito Yamazaki}
\affiliation{Kavli Institute for the Physics and Mathematics of the Universe (WPI), \\
 University of Tokyo,  Kashiwa, Chiba 277-8583, Japan}
\author{Kazuya Yonekura}
\affiliation{ Department of Physics, Tohoku University, Sendai 980-8578, Japan
}

\begin{abstract} 
We propose a mechanism for confinement: analytic continuation beyond infinite coupling in the space of the coupling constant. The analytic continuation is realized by renormalization group flows from the weak to the strong coupling regime. We demonstrate this mechanism explicitly for the mass gap in two-dimensional sigma models in the large $N$ limit. Our analysis suggests that the conventional analysis of the  operator product expansion in itself does not necessarily guarantee the existence of a classical solution corresponding to renormalons. We discuss how the renormalon puzzle may be resolved by the analytic continuation beyond infinite coupling.
\end{abstract}

\maketitle


\medskip
\noindent
\textit{Introduction and Setup}---It has been a long-standing problem in high energy physics to analytically prove
the existence of the mass gap for four-dimensional (\4d) pure Yang-Mills theory or QCD.
It is likely that an ultimate solution to this problem requires a novel insight into the 
question: how is the mass gap generated in asymptotically free theories?

In this paper we propose a new mechanism for mass gap generation and for confinement: 
``analytic continuation
beyond infinite coupling constant''. As we will see, this puts together
several interesting ingredients, such as compactification on $\bS^1$, 
renormalization-group (RG) running and analytic continuation of the coupling constant.
We will also comment on implications for the renormalon problem.

The theories we study in this paper are two-dimensional (\2d) non-linear sigma models.
Before discussing that case, however, let us give an example of a similar spirit in a different dimension. 
In the \4d $\CN=2$ super-Yang-Mills theory,
we can sum over quantum corrections~\cite{Nekrasov:2002qd} to obtain the low energy effective coupling.
After resummation of all the corrections, the result can be analytically continued over the moduli space of vacua
from weak to strong coupling regions. 
By perturbing the theory to $\CN=1$,
confinement is triggered by the monopole condensation~\cite{Seiberg:1994rs}.
In the original $\CN=2$ theory, this happens at the point of the moduli space of vacua where the coupling constant becomes infinity. 

We will illustrate our mechanism 
in the case of the \2d $O(N)$-model,
an asymptotic-free sigma model whose target space is the $(N-1)$-dimensional sphere $\bS^{N-1}$. It is straightforward to repeat a similar analysis for the \2d
$\mathbb{CP}^{N-1}$-model. These models are known to have mass gaps, and 
traditionally have been studied as toy models for \4d QCD.
More recently it has been pointed out \cite{Yamazaki:2017ulc,Yamazaki:2017dra} that  the \2d $\mathbb{CP}^{N-1}$-model
arises from a compactification of the \4d pure $SU(N)$ Yang-Mills theory.

Let $\vec{n}=\{ n_i \}_{1 \leq i \leq N}$ be fields with values on the sphere $\bS^{N-1}$ of unit radius ($\vec{n}^2=1$). The Lagrangian
of the $O(N)$-model is given by 
\beq
\CL=\frac{1}{2g^2} \left[ \partial_\mu \vec{n} \partial^\mu \vec{n}+ \alph(\vec{n}^2- 1) \right], \label{eq:onlag2}
\eeq
where $\alph$ is the Lagrange multiplier implementing the constraint $\vec{n}^2=1$.
The operator equations of motion are $ - \partial^2 \vec{n} + \alph \vec{n} =0$ and $\vec{n}^2-1=0$.
The vacuum expectation value (VEV) of $\alph$ is then given by 
$\vev{\alph}= - \vev{\partial_\mu \vec{n}\partial^\mu \vec{n}}$. 
The VEV $\vev{\alph}$ also appears as a mass term for $\vec{n}$, and hence determines 
the mass gap $\Delta$ of the theory as $\Delta = \sqrt{\vev{\alph}}$ up to $1/N$ corrections. The question is then to understand the mechanism for the generation of this gap.

\medskip
\noindent
\textit{Analysis for small $R$}---We will consider this model on spacetime $\bR^1 \times \bS^1$.
Let $\sigma$ be the coordinate of $\bS^1$ with $\sigma \sim \sigma +2\pi R$, and let $\tau$ be the coordinate of $\bR^1$.

If the radius $R$ of $\bS^1$ is much smaller than the dynamical scale of the theory, we 
would expect that the theory is weakly coupled and a power-series (or trans-series) expansion is reliable.
Let us therefore first choose $R$ to be small.

We can solve the model in the large $N$ limit \cite{DAdda:1978vbw,Witten:1978bc}.
By integrating out the fields $\vec{n}$ in \eqref{eq:onlag2}, we get the effective Lagrangian of $\alph$,
\beq
\CL_{\rm eff}(\alph) = \frac{N}{2} \Tr \log (-\partial^2+\alpha) - \frac{1}{2g^2} \alph .
\eeq
For constant $\alph$, we get
\beq
& \Tr \log (-\partial^2+\alph)  \nonumber \\
&\quad =\frac{1}{2\pi R} \sum_{m \in \bZ} \int \frac{d^{d-1}k}{(2\pi)^{d-1}} \log (k^2+R^{-2}m^2 +\alph) \nonumber \\
&\quad =-\frac{2\Gamma(-\frac{d-1}{2})}{(4\pi)^{\frac{d+1}{2}} R}  \sum_{m \in \bZ}(R^{-2}m^2 + \alph)^{\frac{d-1}{2}} ,
\eeq
where we used the dimensional regularization.
The VEV of $\alph$ is determined by the saddle point equation $\partial \CL_{\rm eff}(\alph)/\partial\alph=0$.
Defining (up to $1/N$ corrections)
\beq
\tilde{\Delta}^2 : =R^2\Delta^2 =R^2\vev{\alph} ,
\eeq
the equation $\partial \CL_{\rm eff}(\alph)/\partial \alph=0$ is given by
\beq
  \sum_{m \in \bZ} \frac{1}{(m^2 +\tilde{\Delta}^2 )^{\frac{3-d}{2}}} =\frac{(4\pi)^{\frac{d-1}{2}} R^{d-2}} {2\Gamma(\frac{3-d}{2})} \cdot \frac{1}{\lambd} ,
\eeq
where $\lambd=Ng^2/(4\pi)$
is the 't~Hooft coupling.

Now we need to perform renormalization. Notice that we can rewrite
\beq
  \sum_{m \in \bZ} \frac{1}{(m^2 +\tilde{\Delta}^2 )^{\frac{3-d}{2}}} = F(\tilde{\Delta})+2\zeta(3-d) ,
\eeq
where we defined
\beq
F(\tilde{\Delta})&:= \frac{1}{\tilde{\Delta}^{(3-d)}} +2 \sum_{m = 1}^\infty \left[ \frac{1}{(m^2 +\tilde{\Delta}^2 )^{\frac{3-d}{2}}} -\frac{1}{m^{3-d}}\right] \nonumber \\
&\xrightarrow{d \to 2} \frac{1}{\tilde{\Delta}} +2 \sum_{m = 1}^\infty \left[ \frac{1}{(m^2 +\tilde{\Delta}^2 )^{\frac{1}{2}}} -\frac{1}{m}\right] .
\eeq
Here $\zeta(3-d)$ is the zeta function which has a pole as $\zeta(3-d) \xrightarrow{d \to 2} 1/(2-d) $.
To renormalize it, we set the bare coupling $\lambd$ in terms of a renormalized coupling $\lambd_\mu$ as
\beq
\frac{1}{\lambd} = \mu^{d-2} \left(\frac{1}{\lambd_\mu} + \frac{2}{2-d}+C \right) ,
\eeq
where $\mu$ is a renormalization scale and $C$ is an arbitrary finite constant which specifies the renormalization scheme.
We choose this constant so that we get
\beq
&\frac{(4\pi)^{\frac{d-1}{2}} (R \mu)^{d-2}} {2\Gamma(\frac{3-d}{2})} \cdot \left(\frac{1}{\lambd_\mu} + \frac{2}{2-d}+C \right)  -2\zeta(3-d) \nonumber \\
&\xrightarrow{d \to 2}  \frac{1}{\lambd_\mu} -2\log(R\mu)  = \frac{1}{\lambd_{R^{-1}}} .\label{eq:RGcouling}
\eeq
Therefore, we finally obtain the equation which determines $\tilde{\Delta}$ as
\beq 
\lambd_{R^{-1}}  = \frac{1}{F(\tilde{\Delta})} ,  \label{eq:fundamental}
\eeq
where $\lambd_{R^{-1}}$ is the coupling at the renormalization scale $\mu = R^{-1}$.

The solution of \eqref{eq:fundamental} can be expanded as
\beq
\tilde{\Delta}=\lambd_{R^{-1}}  - \zeta(3) (\lambda_{R^{-1}})^4 + \cdots, 
\eeq
and hence the mass gap at the leading order in $\lambd_{R^{-1}}$ (and in the large $N$ limit) is $\Delta \simeq \lambda_{R^{-1}}/R$. 
We can understand this more directly as follows.

Let us go back to the case of finite $N$.
When $R$ is small, by integrating out the Kaluza-Klein (KK) modes along $\bS^1$, 
we get an effective quantum mechanics whose target space is still $\bS^{N-1}$.

At the leading order, the effect of integrating out the KK modes is to renormalize the coupling constant to $g_{R^{-1}}$,
i.e., the coupling constant at the renormalization scale $\mu = R^{-1}$.
The one-dimensional (\1d) Lagrangian is given by $L_{1d} = (2\pi R /  g^2_{R^{-1}} ) (\partial_\tau \vec{n})^2 /2$
(where the constraint $\vec{n}^2=1$ is implicit), and hence the Hamiltonian is
\beq
H_{1d} =\frac{1}{2} \cdot \frac{g^2_{R^{-1}}}{2\pi R} \cdot \Laplace_{\bS^{N-1}}, \label{eq:hamiltonian}
\eeq
where $\Laplace_{\bS^{N-1}}$ is the Laplacian on $\bS^{N-1}$.
The Laplacian $\Laplace_{\bS^{N-1}}$ has the spectrum $\ell (\ell+N-2)~(\ell=0,1,2,\cdots)$, and hence we obtain the mass gap (or energy gap) $\Delta$ 
between $\ell=0$ and $\ell=1$ as
\beq
 \Delta=\frac{g^2_{R^{-1}}}{4\pi R} (N-1) + \CO(\lambd_{R^{-1}}^2 ) .\label{eq:1stmassgap}
\eeq
This gives $\tilde{\Delta}=R\Delta= \lambd_{R^{-1}} + \CO(\lambd_{R^{-1}}^2,\lambd_{R^{-1}}/N )$ in the large $N$ limit.

\medskip
\noindent
\textit{Analytic Continuation to Large $R$}---We now wish to go to the large $R$ regime.
The \1d quantum-mechanical description given above will then no longer be valid.
However, the large $N$ result \eqref{eq:fundamental} is still valid without any change.

The radius $R$ in \eqref{eq:fundamental} changes the value of the 't Hooft coupling $\lambd_{R^{-1} }$, as 
follows from the RG equation:
\beq
\mu \frac{\partial}{\partial \mu} \left(\frac{1}{\lambd_\mu} \right)=2. \label{eq.RG}
\eeq
This result is exact at the leading order of the large $N$ expansion in the renormalization scheme specified above.
The 't Hooft coupling $\lambd_{R^{-1}}$ is small and positive when $R$ is small, but it becomes large and eventually becomes infinite and negative as the $R$ is increased.
Thus, the RG flow forces $\lambd_{R^{-1}}$ to go ``beyond the infinite coupling" when $R$ is increased.

Now we can discuss our main point. The function $1/F(\tilde{\Delta}) =\tilde{\Delta}+ \zeta(3) \tilde{\Delta}^4+ \cdots$ is an analytic function of $\tilde{\Delta}$
having a finite nonzero radius of convergence (which in this case is $1$ because the nearest singularity appears at $\tilde{\Delta} = \pm i$). 
Thus, near the origin, the inverse function theorem tells us that there is an analytic function $G(\lambd_{R^{-1}})$ 
such that
\beq
\tilde{\Delta}=G(\lambd_{R^{-1}} )=\sum_{k=1}^\infty a_k (\lambd_{R^{-1}} )^k, \label{eq:Gexpand}
\eeq
where $a_1=1$, $a_2=a_3=0$, $a_4 = -\zeta(3)$ and so on.
This expansion also has a finite nonzero radius of convergence.
This means that $\tilde{\Delta}$ has a convergent power series expansion in terms of the coupling constant $\lambd_{R^{-1}}$,

Furthermore, the function $F(\tilde{\Delta})$ is an analytic function on the positive real axis $0 < \tilde{\Delta} < \infty$
and it is monotonic as a function of $\tilde{\Delta}$. Again by the inverse function theorem, this means that $G(\lambd_{R^{-1}} )$ should also be analytic 
in the range $(\lambda_{R^{-1}})^{-1} \in   (-\infty, \infty)$.

In the region $\tilde{\Delta} \to \infty$, $F(\tilde{\Delta})$ behaves as
\beq
F(\tilde{\Delta}) \to -2 (\log \tilde{\Delta} - \log 2 + \gamma) + \CO(\tilde{\Delta}^{-1}),
\eeq
where $\gamma$ is the Euler's constant. By \eqref{eq:fundamental}, we see that $\tilde{\Delta} \to \infty$ corresponds to $(\lambda_{R^{-1}})^{-1} \to -\infty$, and
\beq
G(\lambd_{R^{-1}}) \to 2e^{-\gamma} \exp\left( - \frac{1}{2\lambd_{R^{-1}}}\right). \label{eq:Gcontinued} 
\eeq
By analytic continuation,
we can smoothly go from the region $(\lambda_{R^{-1}})^{-1} \gg 1$ where \eqref{eq:Gexpand} is valid to the region $(\lambda_{R^{-1}})^{-1} \ll -1$ where \eqref{eq:Gcontinued} is valid by increasing $R$ from $R \sim 0$ to $R \to \infty$.
Notice that the better variable in this analytic continuation is not $\lambd_{R^{-1}}$, but $(\lambda_{R^{-1}})^{-1} = -2\log(R) +\text{const}$.
In the region $1/\lambd_{R^{-1}} \ll -1$, we obtain
\beq
\Delta=\frac{\tilde{\Delta}}{R} \to \Lambda,
\eeq
where the dynamical scale $\Lambda$ is defined as
\beq
\Lambda:=2e^{-\gamma}  \mu \exp\left( - \frac{1}{2\lambd_\mu}\right).
\eeq
This is the famous mass gap generation in the $O(N)$ sigma model.

Now it is clear what is going on. When the $R$ is small, the coupling $\lambd_{R^{-1}}$ is small
and the function $G(\lambd_{R^{-1}})$ is an analytic function of $\lambd_{R^{-1}}$ with a finite nonzero radius of convergence. As we increase $R$,
the coupling becomes large, and at some point it becomes infinity. The inverse coupling $1/\lambd_{R^{-1}}$ smoothly changes from positive values
to negative values. Then, for very large $R$, the $1/\lambd_{R^{-1}}$ is negative and has large absolute value, meaning that $\lambd_{R^{-1}}$
is small and negative. The function $G(\lambd_{R^{-1}})$ can be analytically continued in this process. However, the function $G(\lambd_{R^{-1}})$
has branch cuts somewhere outside the real axis, and after the above process of analytic continuation, the
$G(\lambd_{R^{-1}})$ for small negative values of $\lambd_{R^{-1}}$ is in a different Riemann sheet. 
On the first sheet, we have $G(\lambd_{R^{-1}}) \sim \lambd_{R^{-1}}$ for $|\lambd_{R^{-1}}| \ll 1$, and in the second sheet we have
$G(\lambd_{R^{-1}})  \sim 2e^{-\gamma} \exp ( - 1/2\lambd_{R^{-1}})$ for $|\lambd_{R^{-1}}| \ll 1$.
By this analytic continuation, the mass gap $\Delta $ goes from $\lambd_{R^{-1}}/R$ to $\Lambda$.

The explicit forms of $F(\Delta)$ and $G(\lambd_{R^{-1}})$ are complicated,
so it might be helpful to have in mind the following toy functions,
\beq
F_{\rm toy}(\tilde{\Delta}) &= \frac{1}{\log(1+\tilde{\Delta})} - 2\log(1+\tilde{\Delta}), \\
G_{\rm toy}(\lambd_{R^{-1}}) &=\exp \left ( \frac{\sqrt{1+ 8\lambd_{R^{-1}}^2} - 1}{4\lambd_{R^{-1}}} \right) -1,
\eeq 
which capture qualitative features of the actual functions $F(\tilde{\Delta})$ and $G(\lambd_{R^{-1}})$.
Near the origin $|\lambd_{R^{-1}}| \ll 1$, the $G_{\rm toy}$ has the expansion $G_{\rm toy}(\lambd_{R^{-1}}) =\lambd_{R^{-1}}+\cdots$.
However, after passing the point at infinity $\lambd_{R^{-1}} =\infty$, the expression $+\sqrt{1+ 8\lambd_{R^{-1}}^2}/\lambd_{R^{-1}}$ changes into
$-\sqrt{1+ 8\lambd_{R^{-1}}^2}/\lambd_{R^{-1}}$ 
and hence, for $-1/\lambd_{R^{-1}} \gg 1$, we get $G_{\rm toy}(\lambd_{R^{-1}}) \to \exp(-1/2\lambd_{R^{-1}})$.

\medskip
\noindent
\textit{Remark on scheme dependence}---
Our discussion above depended on the renormalization scheme which leads to \eqref{eq:RGcouling}.
Let us comment on the issue of scheme dependence.

We can take the coupling $\lambda$ to be a complex variable, simply because the path integral makes sense for complex $\lambda$,
at least if the real part of $(\lambda)^{-1}$ is positive and large. (For this formal argument to be true, no IR divergences should spoil the path integral.)
We analytically continue physical quantities from the region ${\rm Im} (\lambda^{-1}) \gg 1$ to other values. Then we obtain a complex manifold $\CM$ of the space of 
complexified coupling constants.

In this description, scheme dependence of the coupling constant simply corresponds to the choice of local coordinate systems on $\CM$.
In our example above, we may use the coordinate $\lambd$ near $\lambd=0$ and another coordinate $\lambd'= (\lambd)^{-1}$ near  $\lambd=\infty$.
It can happen that $\CM$ is nontrivial due to branch cuts etc. and a single coordinate system is not enough to describe the entire $\CM$ 
in the sense that the Taylor expansion around each point $p \in \CM$ may have only finite radius of convergence $|\lambd - \lambd_p| < a_p$ for some $a_p$ and a coordinate system $\lambda$. 
However, we can perform analytic continuation beyond that radius as far as we avoid singularities.

For example, for a class of $\CN=2$ superconformal theories,
the space of couplings even before complexification (in the above sense) is already complicated~\cite{Gaiotto:2009we}.
Examples which are closer to our discussion may be found in \cite{Costello:2019tri,CM}.

The RG equation is a flow equation on $\CM$. As we change $R$, the coupling $\lambda_{R^{-1}}$ moves in the space $\CM$.
If one do not want to consider the space $\CM$, one may just fix an RG scale $\mu$ and the coupling $\lambda_\mu$, and perform analytic continuation
in terms of the variable $\log( R)$. This is a scheme independent way of doing the analytic continuation, but the interpretation in terms of the coupling constant is less clear.

\medskip
\noindent
\textit{Implications for Renormalons}---The fact that the analytic continuation works in the discussion above
has interesting implications for the famous renormalon problem~\cite{tHooft:1977xjm,LeGuillou:1990nq} (see e.g.\ \cite{David:1982qv,David:1983gz,David:1985xj,Novikov:1984ac,Beneke:1998eq}
for early works on the renormalons for large $N$ sigma models).

It should be pointed out that renormalons are not seen at the leading order of $1/N$ expansion in the current model. But we expect that the main message below
is still valid in the subleading orders in $1/N$. It would be interesting to study it explicitly at the subleading orders of $1/N$.

Let us briefly recall the conventional argument for renormalons in asymptotically free theories \cite{Parisi:1978az,Weinberg:1996kr}.

Suppose we want to compute some correlation functions on flat Euclidean space $\bR^d$. 
Let $J(x)$ be an operator of the theory.
By the operator product expansion (OPE) we get $J(x) J(0)=C_0(x)+C_1(x) O(0) +\cdots$, 
where $O(x)$ is some operator of the theory. The OPE coefficients $C_0(x), C_1(x),\cdots$ may be computed 
at short distances if $|x|$ is small enough, so ``weak coupling'' computation might be possible for them. 
(This point, however, is actually not so obvious as we discuss later.)
By using the OPE, the correlation function is given as
\beq
\vev{J(x) J(0)} = C_0(x)+C_1(x) \vev{ O}+\cdots.
\eeq
By dimensional counting, the VEV $\vev{O}$ is expected
to be of order $\Lambda^{D_O}$, where $\Lambda$ is the strong coupling dynamical scale and $D_O$ is the mass dimension of $O$ (possibly including the anomalous dimension).
This $\Lambda$ is given as
\beq
\Lambda= \mu \exp \left(-\int^{\lambd_\mu} \frac{d \lambd}{\beta(\lambda)}\right) = \mu\exp\left(-\frac{1}{\beta_0 \lambd_\mu } +\cdots \right) ,\label{eq:dyn}
\eeq
where $\lambda_\mu \propto g_\mu^2$ is the coupling squared at the renormalization scale $\mu$, and 
$\beta= \mu(\partial \lambd/\partial \mu) = - \beta_0 \lambda^2 +\cdots$ is the beta function.
In the case of the $O(N)$ sigma model in the large $N$ limit discussed above, we have $\beta_0=2$ for $\lambda= N g^2/4\pi$.

If the correlation function contains a term proportional to $\vev{O} \sim \Lambda^{D_O} \sim \exp( - D_O/\beta_0 \lambd_\mu )$,
it is often said that there is a pole in the Borel plane.
Moreover, it is sometimes claimed that there exists a corresponding 
solution of the classical equations of motion (i.e.\ a saddle point of the path integral) with the value of the action given by $S = D_O/\beta_0 \lambd_\mu $.
This question motivated many studies, including two-dimensional sigma models compactified on $\bS^1$. 
See e.g.\ \cite{Bruckmann:2007zh,Brendel:2009mp,Dunne:2012ae,Dunne:2012zk,Fujimori:2018kqp,Bruckmann:2019mky,Ishikawa:2019tnw} for detailed studies in twisted $\bS^1$ compactification of the $\mathbb{CP}^{N-1}$ sigma model. Although saddle points of the classical action do exist~\cite{Dunne:2012ae,Fujimori:2018kqp} in such twisted compactification, there are some evidence~\cite{Ishikawa:2019tnw,Ishikawa:2019oga} 
(see also the original work \cite{Anber:2014sda} in \4d) that those classical solutions do not correspond to renormalons.

Our analysis suggests that the existence of saddle points of the classical action is not necessary for a solution to the renormalon problem.
As stressed before, 
the mass gap $\Delta$ is a convergent power series of $\lambda_{R^{-1}}$
with finite radius of convergence.
Moreover, $\tilde{\Delta}^2/R^2 = \vev{\alph} $ and $\vev{\alph} =  - \vev{\partial_\mu \vec{n}\partial^\mu \vec{n}}$
by equations of motion as mentioned before.
Therefore, the VEV of the operator $O = \partial_\mu \vec{n}\partial^\mu \vec{n}$ is produced 
by a convergent power series in the weak coupling regime.
Instead of a classical solution corresponding to renormalons, the scale $\Lambda^{D_O}$
is generated by an honest analytic continuation beyond infinite coupling. 
We conjecture that this is also the case for each subleading order of the $1/N$ expansion.

We expect that this is also the case in any theory for generic operators $O$ (which are not protected by symmetries).
What generates $ \vev{O}  \xrightarrow{R \to \infty} \Lambda^{D_O} \sim \mu^{D_O} \exp( - D_O/\beta_0 \lambd_\mu )$ is not a classical saddle point, 
but the analytic continuation beyond infinite coupling $\lambd_{R^{-1}}=\infty$
to go to the different Riemann sheet on the complex plane of the coupling constant $\lambd_{R^{-1}}$.

It should be stressed that our argument does not preclude the existence of 
saddle points corresponding to singularities on Borel planes: Borel planes do contain singularities in general. 
For example, in theories with finite $N$, or even in theories with large $N$ but with twisted compactification,
we expect to have singularities on the Borel plane corresponding to classical saddle points.
In these cases, the weak coupling expansion of $\vev{O}$ is given by a trans-series rather than an expansion with a finite radius of convergence.
To make the trans-series expansion possible, an appropriate compactification should be done so that the theory is weakly coupled and there is no issue of infrared divergences.
(The authors' point of view on this issue is stressed in \cite{Yamazaki:2017ulc}. See also \cite{Itou:2018wkm} for a lattice simulation in the setup of \cite{Yamazaki:2017ulc}.) Then we may rely on resurgence theory (see \cite{Marino:2012zq,Aniceto:2018bis} for reviews) to obtain a complete analytic function of the coupling, and perform the analytic continuation beyond infinity.

\medskip
\noindent
\textit{Factorial growth and Renormalons}---Even though the existence of a classical saddle point with the value of the action $S= D_O/\beta_0 \lambd $
is not guaranteed as discussed above, a factorial growth of power series of the form
\beq
\sum n! ( \beta_0 \lambda/D)^n \label{eq:factorial0}
\eeq
is clearly observed in the literature, where $D$ is an integer. 
See \cite{Beneke:1998ui} for a review of early development,
and \cite{Bauer:2011ws,Bali:2014fea,Marino:2019eym} for explicit verification of the existence of such factorial growth.
Let us comment on how such a series may be obtained.

Let us consider an integral defined by
\beq
I_D(Q) = Q^{-D} \int_{Q^{-1}}^\infty dR\,  R^{-D} \Delta(R) 
\eeq
where $\Delta(R) = R^{-1} G(\lambd_{R^{-1}})$ is the mass gap as a function of $R$,
and $Q$ is a variable with mass dimension 1.
This is meant to be a toy integral to illustrate our point, and we do not ask for its physical meaning.

The first point to be emphasized is that this integral is completely well-defined
as long as $D > 1$. The mass gap $\Delta(R)$ is well-defined by the analytic continuation beyond infinity,
and it behaves as $\Delta(R) \to \Lambda$ as $R \to \infty$.
Therefore the integral is convergent for $D>1$.

Moreover, the integral has a formal expansion in terms of the coupling $\lambda_Q $ at the RG scale $\mu = Q$.
By changing the integration variable as $R = Q^{-1} \exp( t )$,
the integral can be written as
\beq
I_D(Q) &= \int_0^\infty dt\, e^{ - D t } \, G(\lambd_{Q \exp(-t) })  \nonumber \\
&=  \int_0^\infty dt\, e^{ - D t } \, G\left( \frac{\lambd_{Q }}{ 1 - \beta_0 \lambd_Q  t } \right),
\eeq
where $\beta_0=2$ is the coefficient of the one-loop beta function of $\lambda$ (recall \eqref{eq.RG}). 
We can expand $G$ as in \eqref{eq:Gexpand}, and then expand $ \lambd_{Q }/ (1 - \beta_0 \lambd_Q  t ) $
in terms of $\lambda_Q$. 

Let us keep only the leading term in the $G$-expansion,
but keep all orders of the expansion of $ \lambd_{Q }/ (1 - \beta_0 \lambd_Q  t ) $ which is related to the RG flow.
Because $G(\lambd_{R^{-1}} ) =  a_1\lambd_{R^{-1}}  + a_2 \lambd_{R^{-1}}^2 +  \cdots$ (with $a_1=1)$, we get
\beq
I_D(Q) & \supset \int_0^\infty dt\, e^{ - D t }  \sum_{n=0}^\infty a_1 \lambd_Q (\beta_0 \lambd_Q t)^n  \nonumber \\
& = \frac{a_1 \lambd_Q}{D} \sum_{n=0}^\infty  n! \left( \frac{\beta_0 \lambd_Q}{D} \right)^n.
\eeq
This is exactly of the form \eqref{eq:factorial0}---the formal expansion of the completely well-defined integral $I_D(Q)$
gives the factorial growth.

The integral $I_D(Q)$ above was chosen to imitate (in a simplified way) the integral which appears in a correlation function
$\vev{J(x) J(0)} = C_0(x) +\cdots$ in QCD which is known to have a factorial growth. (See \cite{Beneke:1998ui} for a review.)
There, the integration variable is a momentum rather than $R$, but the structure of the integral is similar to that of $I_D(Q)$ if
we focus on the infrared (IR) region (i.e.\ the IR renormalon). Roughly, summing up bubble diagrams of the gluon propagator 
corresponds to taking the RG improved coupling $\lambd_Q /(1 - \beta_0 \lambd_Q t)$, and neglecting
other diagrams may correspond to neglecting higher order terms in the expansion of $G(\lambda_{R^{-1}})$.

What makes the integral $I_D(Q)$ well-defined is the analytic continuation of the integrand $G(\lambda_{R^{-1}})$ beyond infinite coupling.
We conjecture that this is how the factorial growth should be understood in the actual renormalon problem:
the factorial growth of $I_D(Q)$ occurs because the integral involves the strongly coupled region $R \to \infty$.
In a completely weakly coupled setup without any IR divergences (such as \cite{Yamazaki:2017ulc}), we claim that there is no issue of the factorial growth
associated to IR renormalons.

\medskip
\noindent
\textit{Acknowledgements}---The authors thank H.~Suzuki and Y.~Sumino for helpful conversation.
This work is supported in part by JSPS KAKENHI Grant-in-Aid (MY: 17KK0087, 19K03820, 19H00689
and KY: 17K14265).

\bibliography{ref}
  
\end{document}